\documentclass[pra,twocolumn,floatfix,showpacs,superscriptaddress]{revtex4}

\usepackage{amsfonts}
\usepackage{times}
\usepackage{graphicx}

\begin{document}

\title{Abelian and non-Abelian geometric phases in adiabatic open quantum
systems}
\author{M. S. Sarandy}
\affiliation{Instituto de F\'{\i}sica de S\~ao Carlos, Universidade de S\~ao Paulo, S\~ao
Carlos, SP, 13560-970, Brazil.}
\author{D. A. Lidar}
\affiliation{Departments of Chemistry, Electrical Engineering-Systems, and Physics,
University of Southern California, Los Angeles, CA 90089, USA.}

\begin{abstract}
We introduce a self-consistent framework for the analysis of both Abelian
and non-Abelian geometric phases associated with open quantum systems,
undergoing cyclic adiabatic evolution. We derive a general expression for
geometric phases, based on an adiabatic approximation developed within an
inherently open-systems approach. This expression provides a natural
generalization of the analogous one for closed quantum systems, and we prove
that it satisfies all the properties one might expect of a good definition
of a geometric phase, including gauge invariance. A striking consequence is
the emergence of a finite time interval for the observation of geometric
phases. The formalism is illustrated via the canonical example of a spin-1/2
particle in a time-dependent magnetic field. Remarkably, the geometric phase
in this case is immune to dephasing and spontaneous emission in the
renormalized Hamiltonian eigenstate basis. This result positively impacts
holonomic quantum computing.
\end{abstract}

\pacs{03.65.Vf, 03.65.Yz, 03.67.-a, 03.65.Ta}
\maketitle

\section{Introduction}

More than twenty years ago, Berry first observed that quantum systems may
retain a memory of their motion in Hilbert space through the acquisition of
geometric phases~\cite{Berry:84}. Remarkably, these phase factors depend
only on the geometry of the path traversed by the system during its
evolution. Soon after Berry's discovery, geometric phases became a subject
of intense theoretical and experimental studies~\cite{GP-Book:89}. In recent
years, renewed interest has arisen in the study of geometric phases in
connection with quantum information processing~\cite{Zanardi:99,Jones:00}.
Indeed, geometric, or holonomic quantum computation (QC) may be useful in
achieving fault tolerance~\cite{Steane:03}, since the geometric character of
the phase provides protection against certain classes of errors~\cite%
{Solinas:04,Zhu:05,Guridi:05}. However, a comprehensive investigation in
this direction requires a generalization of the concept of geometric phases
to the domain of \emph{open} quantum systems, i.e., quantum systems which
may decohere due to their interaction with an external environment.

Geometric phases in open systems, and more recently their applications in
holonomic QC, have been considered in a number of works, since the late
1980's. The first approach to the subject used the Schr\"{o}dinger equation
with non-Hermitian Hamiltonians~\cite{Garrison:88,Dattoli:90}. This is a
phenomenological, non-rigorous approach (e.g., it cannot guarantee
completely positivity). A consistent non-Hermitian Hamiltonian description
of an open system in general requires the theory of stochastic Schr\"{o}%
dinger equations \cite{Gardiner:book}. Nevertheless, this approach for the
first time indicated that complex Abelian geometric phases should appear for
systems undergoing cyclic evolution. In Refs.~\cite%
{Ellinas:89,Gamliel:89,Romero:02,Whitney:03,Whitney:05,Kamleitner:04},
geometric phases acquired by the density operator were analyzed for various
explicit models within a master equation approach, but no general theory was
formulated for open system geometric phases. In Refs.~\cite%
{Carollo:03,Guridi:05}, the quantum jumps method was employed to provide a
definition of geometric phases in Markovian open systems (related
difficulties with stochastic unravellings have been pointed out in Ref.~\cite%
{Bassi:05}). In another approach the density operator, expressed in its
eigenbasis, was lifted to a purified state \cite{Tong:04,Rezakhani:05}. In
Ref.~\cite{Marzlin:04}, a formalism in terms of mean values of distributions
was presented. An interferometric approach for evaluating geometric phases
for mixed states evolving unitarily was introduced in Ref.~\cite{Vedral:00}
and extended to non-unitary evolution in Refs.~\cite{Faria:03,Ericsson:03}.
This interferometric approach can also be considered from a purification
point of view~\cite{Vedral:00,Ericsson:03}. This multitide of different
proposals have revealed various interesting facets of the problem.
Nevertheless, the concept of adiabatic geometric phases in open systems
remains unresolved in general, since most of the previous treatments did not
employ an adiabatic approximation genuinely developed for open
systems. Note that the applicability of
the closed systems adiabatic approximation~\cite{Messiah:book} to open
systems problems is not a priori clear and should be justified on a
case-by-case basis. Moreover, almost all of the previous works on open
systems geometric phases were concerned with the Abelian (Berry phase) case.
Exceptions are the very recent Refs.~\cite{Guridi:05,Florio:06,Trullo:06},  
which discuss both non-adiabatic and adiabatic dynamics, but employ
the standard adiabatic theorem for closed systems in the latter
case.

In this work, we introduce a self-consistent open-systems framework, based
on a recent generalization of the adiabatic approximation~\cite%
{SarandyLidar:04}, which allows for a general definition and evaluation of
both Abelian and non-Abelian geometric phases in open systems undergoing
cyclic adiabatic evolution. As we shall show, this approach yields new
insights and lends itself to a simple and elegant generalization of the
concept of geometric phases. An important feature emerging from this picture
is the appearance of a distinguished time-scale for the observation of
adiabatic geometric phases in open systems. We illustrate our results by
considering the canonical example of a spin-$1/2$ in a magnetic field. In
this example, we find a remarkable robustness of the geometric phase against
both dephasing and spontaneous emission in the instantaneous renormalized
Hamiltonian eigenstate basis. This result should have a positive impact on
the robustness against external disturbances of holonomic QC.

We note that an alternative theory for adiabaticity in open systems
was recently developed by
Thunstr\"{om}, {\AA }berg and Sj\"oqvist, for systems coupled
weakly to their environment. This theory was then employed to
study a non-Abelian geometric phase gate in holonomic QC
\cite{Thunstrom:05}. The main difference between the approach of
Ref.~\cite{Thunstrom:05} and our approach to adiabaticity in open
systems, is that Thunstr\"{om} et al. emphasize the decoupling of
eigenspaces of the system Hamiltonian, while we focus on the entire
superoperator. Central conclusions, such as the breakdown of
adiabaticity in open systems, are shared by the two approaches. A full
comparison is beyond the scope of the present work.

The structure of this paper is as follows. In the next section we
review the coherence vector approach to solving quantum master
equations, the Jordan form of the superoperator, and the notion of
adiabaticity in open quantum systems. In Section~\ref{GP} we derive
the geometric phase in open systems, in both the Abelian (Berry phase)
and non-Abelian cases. We prove that this geometric phase satisfies
the expected properties, such as a proper closed system limit and
gauge invariance. In Section~\ref{apps} we focus on applications,
namely, we show that our theory predicts that there is a distinguished
time-scale for open system geometric phases (related to the breakdown
of adiabaticity), and then consider the example of a spin-$1/2$
coupled to a slowly varying magnetic field, in the presence of
dephasing and spontaneous emission. We conclude in Section~\ref{conc}.

\section{Master equations and the adiabatic regime of open quantum systems}

\label{master}

Open quantum systems typically do not undergo unitary dynamics, i.e., they
are not governed by the Schr\"{o}dinger equation, or even by its
non-Hermitian generalization~\cite{Garrison:88,Dattoli:90}. Instead, quite
generally we may consider open quantum systems evolving under a
convolutionless master equation~\cite{Breuer:book} 
\begin{equation}
{\partial {\rho }}/\partial t=\mathcal{L}[\vec{R}(t)]\rho (t),
\label{eq:t-Lind1}
\end{equation}
where $\mathcal{L}$ is a superoperator which depends on time only through a
set of parameters $\vec{R}(t)\equiv \vec{R}$. The Lindblad equation~\cite%
{Lindblad:76} is an important example of this class of master equations: 
\begin{equation}
{\partial {\rho }}/\partial t=-i\left[ H,\rho \right] +\frac{1}{2}
\sum_{i}\left( [\Gamma _{i},\rho \Gamma _{i}^{\dagger }]+[\Gamma _{i}\rho
,\Gamma _{i}^{\dagger }]\right) ,  \label{eq:t-Lind2}
\end{equation}
where we have suppressed the explicit dependence of the operators on $\vec{R}%
(t)$. Here $H$ is the effective Hamiltonian of the open system (it is
renormalized, i.e., contains the \textquotedblleft Lamb
shift\textquotedblright\ -- the unitary contribution of the system-bath
interaction \cite{Alicki:87,Lidar:CP01}), the $\Gamma _{i}$ are operators describing the system-bath
interaction, and we work in $\hbar =1$ units. In this work we consider the
general class of convolutionless master equations~(\ref{eq:t-Lind1}), and in
a later section illustrate our formalism with an example using the case of
Eq.~(\ref{eq:t-Lind2}). In this example, of a spin-$1/2$ in a magnetic
field, we allow \emph{both} $H$ and the $\Gamma _{i}$ to depend on $\vec{R}%
(t)$. In a slight abuse of nomenclature, we will refer to the implicitly
time-dependent generator $\mathcal{L}$ [Eq.~(\ref{eq:t-Lind1})] as the
Lindblad superoperator and the $\Gamma _{i}$ [Eq.~(\ref{eq:t-Lind2})] as
Lindblad operators. This terminology is usually associated with
time-independent generators~\cite{Lindblad:76,Alicki:87}, but recent work
has clarified the conditions under which Eq.~(\ref{eq:t-Lind2}) with
time-dependent $\Gamma _{i}$ can be derived in the usual Davies \cite%
{Davies:78} weak-coupling limit \cite{ALZ:05}. What is important to note is
that the microscopic weak-coupling limit derivation is consistent with the
general class of master equations postulated here, wherein the Lindblad
operators depend implicitly on time through their explicit dependence on
external control fields. More specifically, in the microscopic derivation
one shows that the Lindblad operators are the Fourier components of the
time-dependent system operator terms in the system-bath interaction
Hamiltonian, where the time-dependence arises by working in the interaction
picture with respect to the \emph{renormalized} system Hamiltonian \cite%
{ALZ:05}. This is how the $\Gamma _{i}$'s appearing here and in our example
below must be interpreted.

In the superoperator formalism, the density matrix for a quantum state in a 
$D$-dimensional Hilbert space is represented by a $D^{2}$-dimensional
\textquotedblleft coherence vector\textquotedblright\ $|\rho \rangle \rangle
=\left( \rho _{1},\rho _{2},\cdots ,\rho _{D^{2}}\right) ^{t}$ and the
Lindblad superoperator $\mathcal{L}$ becomes a $D^{2}\times D^{2}$%
-dimensional supermatrix \cite{Alicki:87}, so that the master equation (\ref%
{eq:t-Lind1}) can be written as linear vector equation in $D^{2}$%
-dimensional Hilbert-Schmidt space, in the form 
\begin{equation}
\partial |\rho \rangle \rangle /\partial t=\mathcal{L}[\vec{R}(t)]|\rho
\rangle \rangle .  \label{eq:Lind3}
\end{equation}
Such a representation can be generated, e.g., by introducing a basis of
Hermitian, trace-orthogonal, and traceless operators [e.g., the $D$%
-dimensional irreducible representation of the generators of su($D$)],
whence the $\rho _{i}$ are the expansion coefficients of $\rho $ in this
basis \cite{Alicki:87}, with $\rho _{1}$ the coefficient of $I$ (the
identity matrix). In this case, the condition $\mathrm{Tr}\rho ^{2}\leq 1$
corresponds to $\left\Vert |\rho \rangle \rangle \right\Vert \leq 1$, $\rho
=\rho ^{\dag }$ to $\rho _{i}=\rho _{i}^{\ast }$, and positive
semidefiniteness of $\rho $ is expressed in terms of inequalities satisfied
by certain Casimir invariants [e.g., of su($D$)] \cite%
{byrd:062322,Kimura:2003-1}. A\ simple and well-known example of this
procedure is the representation of the density operator of a two-level
system (qubit) on the Bloch sphere, via $\rho =(I_{2}+\overrightarrow{v}%
\cdot \overrightarrow{\sigma })/2$, where $\overrightarrow{\sigma }=(\sigma
_{x},\sigma _{y},\sigma _{z})$ is the vector of Pauli matrices [generators
of su($2$)] and $I_{2}$ is the $2\times 2$ identity matrix.

The master equation~generates a non-unitary evolution since $\mathcal{L}$ is
non-Hermitian. In fact, $\mathcal{L}$ need not even be a normal operator ($%
\mathcal{L}^{\dag }\mathcal{L\neq LL}^{\dag }$). Therefore $\mathcal{L}$ is
generally not diagonalizable, i.e., it does not possess a complete set of
linearly independent eigenvectors. Equivalently, it cannot be put into
diagonal form via a similarity transformation. However, one can always apply
a similarity transformation to $\mathcal{L}$ which puts it into the
(block-diagonal) Jordan canonical form~\cite{Horn:book}, namely, 
$\mathcal{L}_{\mathrm{J}}=S^{-1}\mathcal{L}S$. The Jordan form $%
\mathcal{L}_{\mathrm{J}}$ of a $D^{2}\times D^{2}$ matrix $\mathcal{L}$ is a
direct sum of blocks of the form $\mathcal{L}_{\mathrm{J}}=\oplus _{\alpha
=1}^{m}J_{\alpha }$ ($\alpha $ enumerates Jordan blocks), where $m\leq D^{2}$
is the number of linearly independent eigenvectors of $\mathcal{L}$, $%
\sum_{\alpha =1}^{m}n_{\alpha }=D^{2}$ where $n_{\alpha }\equiv \dim
J_{\alpha }$ is the dimension of the $\alpha$th Jordan block, and $J_{\alpha }=\lambda
_{\alpha }I_{n_{\alpha }}+K_{a}$ where $\lambda _{\alpha }$ is the $\alpha $%
th (generally complex-valued) Lindblad-Jordan (LJ) eigenvalue of $\mathcal{L}
$ (obtained as roots of the characteristic polynomial), $I_{n_{\alpha }}$ is
the $n_{\alpha }\times n_{\alpha }$ dimensional identity matrix, and $K_{a}$
is a nilpotent matrix with elements $(K_{a})_{ij}=\delta _{i,j-1}$ ($%
1$'s above the main diagonal), where $\delta $ is the Kronecker symbol.
Instantaneous right $\{|\mathcal{D}_{\beta }^{(j)}[\vec{R}(t)]\rangle
\rangle \}$ and left $\{\langle \langle \mathcal{E}_{\alpha }^{(i)}[\vec{R}%
(t)]|\}$ bi-orthonormal bases in Hilbert-Schmidt space can always be systematically 
constructed such that they obey the orthonormality condition $\langle \langle \mathcal{E}%
_{\alpha }^{(i)}|\mathcal{D}_{\beta }^{(j)}\rangle \rangle =\delta _{\alpha
\beta }\delta ^{ij}$~\cite{SarandyLidar:04}. Here superscripts enumerate basis states
inside a given Jordan block ($i,j\in \{0,...,n_{\alpha }-1\}$).
When ${\cal L}$ is diagonalizable, $\{|\mathcal{D}_{\beta }^{(j)}[\vec{R}(t)]\rangle
\rangle \}$ and $\{\langle \langle \mathcal{E}_{\alpha }^{(i)}[\vec{R}%
(t)]|\}$ are simply the bases of right and left eigenvectors of ${\cal L}$, respectively. 
If ${\cal L}$ is not diagonalizable, these right and left bases can be constructed 
by suitably completing the set of right and left eigenvectors of ${\cal L}$ (See 
Ref.~\cite{SarandyLidar:04} and also Appendix~\ref{appA} for a detailed discussion of the 
left and right basis vectors, including their completeness relation). 
Based on the above considerations we gave, in Ref.~\cite{SarandyLidar:04}, a
definition of adiabaticity in open quantum systems:
\begin{quote}
An open quantum system is said to undergo adiabatic dynamics when its
Hilbert-Schmidt space can be decomposed into decoupled LJ-eigenspaces with
distinct, time-continuous, and non-crossing instantaneous eigenvalues of $%
\mathcal{L}$.
\end{quote}
Note that the key to establishing the concept of adiabaticity in open
systems is to replace the idea of decoupling of the Hamiltonian eigenstates
by that of decoupling of the Jordan blocks of the Lindblad superoperator 
(each Jordan block is associated with an independent eigenstate of ${\cal L}$). 
The definition of adiabaticity given above
implies a condition on the total evolution time $T$ which generalizes the
well-known closed-systems condition, $T\gg \max_{0\leq s\leq 1}|\langle k(s)|%
\frac{dH(s)}{ds}|m(s)\rangle |/|g_{mk}(s)|^{2}$, where $s=t/T$ is the
normalized time, $H$ is the time-dependent Hamiltonian, $|k\rangle $ and $%
|m\rangle $ are eigenstates of $H$, and $g_{mk}$ is the energy gap between
these two states. For further details we refer the reader to Refs. \cite%
{SarandyLidar:04,SarandyLidar:05}; the condition in the case of $1$%
-dimensional Jordan blocks is given in Eq.~(\ref{tcross}) below.
The theory developed in Ref.~\cite{SarandyLidar:04} applies also in the more
general case of explicitly time-dependent generators $\mathcal{L}(t)$, but
since we focus here on geometric phases, we shall only consider implicit
time-dependence as in $\mathcal{L}[\vec{R}(t)]$.

\section{Geometric phases for open systems in cyclic adiabatic evolution}
\label{GP}
  
In order to define geometric phases in open systems, we expand the coherence
vector in the instantaneous right vector basis $\{|{%
\mathcal{D}_{\beta }^{(j)}[\vec{R}(t)]\rangle \rangle }\}$ as 
\begin{equation}
|\rho (t)\rangle \rangle =\sum_{\beta =1}^{m}\sum_{j=0}^{n_{\beta
}-1}p_{\beta }^{(j)}(t)\,e^{\int_{0}^{t}\lambda _{\beta }(t^{\prime
})dt^{\prime }}\,|\mathcal{D}_{\beta }^{(j)}[\vec{R}(t)]\rangle \rangle ,
\label{rho_supop}
\end{equation}
where we have explicitly factored out the dynamical phase $\exp
[\int_{0}^{t}\lambda _{\beta }(t^{\prime })dt^{\prime }]$. The coefficients $%
\{p_{\beta }^{(j)}(t)\}$ play the role of \textquotedblleft
geometric\textquotedblright\ (non-dynamical) amplitudes. We assume that the
open system is in the adiabatic regime, i.e.,
\emph{Jordan blocks associated to
distinct eigenvalues evolve in a decoupled manner} (recall the
definition of open sytems adiabaticity given above).
Then, Eqs.~(\ref{eq:Lind3}),(\ref{rho_supop}) together yield 
\begin{equation}
{\dot{p}}_{\alpha }^{(i)}\,=\,p_{\alpha }^{(i+1)}-\sum_{\beta \,|\,\lambda
_{\beta }=\lambda _{\alpha }}\sum_{j=0}^{n_{\beta }-1}p_{\beta
}^{(j)}\langle \langle \mathcal{E}_{\alpha }^{(i)}|{\dot{\mathcal{D}}}%
_{\beta }^{(j)}\rangle \rangle .  \label{pgen}
\end{equation}
A condition on the total evolution time, which allows for the neglect
of coupling between Jordan blocks used in deriving Eq.~(\ref{pgen}), was given in
Ref.~\cite{SarandyLidar:04}. It is reproduced for the case of
1-dimensional Jordan blocks in Eq.~(\ref{tcross}) below.
Note that, due to the restriction $\lambda _{\beta }=\lambda _{\alpha }$,
the dynamical phase has disappeared. For closed systems, Abelian geometric
phases are associated with non-degenerate levels of the Hamiltonian, while
non-Abelian phases appear in the case of degeneracy. In the latter case, a
subspace of the Hilbert space acquires a geometric phase which is given by a
matrix rather than a scalar. Here, for open systems, one-dimensional Jordan
blocks can be associated either with Abelian or non-Abelian geometric phases
(depending on the possibility of degeneracy) while multi-dimensional Jordan
blocks are naturally tied to a non-Abelian phase.

\subsection{The Abelian Case: Generalized Berry Phase}

Consider the simple case of a non-degenerate one-dimensional Jordan block (a
block that that is a $1\times 1$ submatrix containing an 
eigenvalue of ${\cal L}$). In this case, the
absence of degeneracy implies in Eq.~(\ref{pgen}) that $\lambda _{\beta
}=\lambda _{\alpha }\Rightarrow \alpha =\beta $ (non-degenerate blocks).
Moreover, since the blocks are assumed to be one-dimensional we have $%
n_{\alpha }=1$, which allows us to remove the upper indices in Eq.~(\ref%
{pgen}), resulting in ${\dot{p}}_{\alpha }=-p_{\alpha }\langle \langle 
\mathcal{E}_{\alpha }|{\dot{\mathcal{D}}}_{\alpha }\rangle \rangle $. The
solution of this equation is $p_{\alpha }(t)=p_{\alpha }(0)\exp {[i\gamma
_{\alpha }(t)]}$, with $\gamma _{\alpha }(t)=i\int_{0}^{t}\langle \langle 
\mathcal{E}_{\alpha }(t^{\prime })|{\dot{\mathcal{D}}}_{\alpha }(t^{\prime
})\rangle \rangle dt^{\prime }$. In order to establish the geometric
character of $\gamma _{\alpha }(t)$ we now recall that $\mathcal{L}$ depends
on time implicitly through the parameters $\vec{R}(t)$. Then, for a cyclic
evolution in parameter space along a closed curve $C$, we obtain that the
Abelian geometric phase associated with the Jordan block $\alpha $ is given
by 
\begin{equation}
\gamma _{\alpha }(C)=i\oint_{C}\langle \langle \mathcal{E}_{\alpha }(\vec{R}%
)|\vec{\bigtriangledown}|{\mathcal{D}}_{\alpha }(\vec{R})\rangle \rangle
\cdot d\vec{R}.  \label{gp_abelian}
\end{equation}%
This elegant generalized expression for the geometric phase, which bears
similarity to the original Berry formula~\cite{Berry:84}, is our first main
result. As expected for open systems, $\gamma _{\alpha }(C)$ is complex,
since $\langle \langle \mathcal{E}_{\alpha }|$ and $|{\mathcal{D}}_{\alpha
}\rangle \rangle $ are not related by transpose conjugation. Thus, the
geometric phase may have real and imaginary contributions, the latter
affecting the visibility of the phase.

In Refs. \cite{Garrison:88,Dattoli:90} Garrison and Wright, and Dattoli et
al., found an expression for the open-systems Berry phase that resembles our
Eq.~(\ref{gp_abelian}). Their result is%
\begin{equation}
\tilde{\gamma}_{\alpha }(C)=i\oint_{C}\langle \theta _{\alpha }(\vec{R})| 
\vec{\bigtriangledown}|\psi _{\alpha }(\vec{R})\rangle \cdot d\vec{R}.
\label{eq:GW}
\end{equation}
Here $\{|\psi _{\alpha }\rangle \}$ and $\{|\theta _{\alpha }\rangle \}$ are
a bi-orthonormal set of eigenvectors of a non-Hermitian Hamiltonian $H$ and
its Hermitian conjugate $H^{\dag }$, respectively. There are some important
methodological and technical differences between this and our result. First,
here, instead of working with a phenomenological non-Hermitian Hamiltonian,
we started from the outset with a fully consistent master equation approach,
where the left $\{\langle \langle \mathcal{E}_{\alpha }|\}$ and right $\{|{%
\mathcal{D}}_{\alpha }\rangle \rangle \}$ basis vectors are associated with
the dynamical superoperator $\mathcal{L}$, rather than with the
non-Hermitian Hamiltonian. Second, as a result in our case, the basis vectors
span the $D^{2}$-dimensional Hilbert-Schmidt space, whereas in non-Hermitian
Hamiltonian case the geometric phase expression involves vectors in the
usual $D$-dimensional Hilbert space. As we note below, this implies that in
our case $\gamma _{\alpha }$ is a relative, not absolute, geometric
phase, and hence there is in general no connection between the
expressions (\ref{gp_abelian}) and (\ref{eq:GW}).
Third, unlike Refs. \cite{Garrison:88,Dattoli:90}, where adiabaticity is
imported from the theory of closed systems, we work within a consistent
theory of adiabaticity for open systems, as formulated in Ref.~\cite%
{SarandyLidar:04}.

The expression for $\gamma _{\alpha }(C)$ exhibits a number of important
properties expected from a good definition of a geometric phase:

\begin{itemize}
\item \emph{Geometric character}: $\gamma _{\alpha }(C)$ is geometric, i.e.,
it depends only on the path traversed in parameter space.

\item \emph{Gauge invariance}: $\gamma _{\alpha }(C)$ is gauge invariant,
i.e., we cannot modify (or eliminate) the geometric phase by redefining $%
\langle \langle \mathcal{E}_{\alpha }|$ or $|{\mathcal{D}}_{\alpha }\rangle
\rangle $ by multiplying one of them by a complex factor $\chi (\vec{R})\exp 
{[i\nu (\vec{R})]}$. Indeed, let us define $|{\mathcal{D}}_{\alpha }^{\prime
}\rangle \rangle =\chi \exp {(i\nu )}|{\mathcal{D}}_{\alpha }\rangle \rangle 
$ ($\chi (\vec{R})\neq 0$ $\forall \vec{R}$). Redefinition of right-vectors
automatically implies redefinition of left-vectors due to the normalization
constraint $\langle \langle \mathcal{E}_{\alpha }|{\mathcal{D}}_{\alpha
}\rangle \rangle =1$, so that $\langle \langle \mathcal{E}_{\alpha }^{\prime
}|=\langle \langle \mathcal{E}_{\alpha }|\chi ^{-1}\exp {(-i\nu )}$.
Therefore, $\langle \langle \mathcal{E}_{\alpha }^{\prime }|\vec{%
\bigtriangledown}|{\mathcal{D}}_{\alpha }^{\prime }\rangle \rangle =\langle
\langle \mathcal{E}_{\alpha }|\vec{\bigtriangledown}|{\mathcal{D}}_{\alpha
}\rangle \rangle +(\vec{\bigtriangledown}\chi )/\chi +i\vec{\bigtriangledown}%
\nu $. Gauge invariance then follows from the computation of $\gamma
_{\alpha }^{\prime }$ using Eq.~(\ref{gp_abelian}), with Stokes's theorem
leading to $\gamma _{\alpha }^{\prime }(C)=\gamma _{\alpha }(C)$. Below we
provide a detailed proof of gauge invariance in the non-Abelian case, which
includes the latter as a special case.

\item \emph{Closed system limit}: if the interaction with the bath vanishes, 
$\gamma _{\alpha }(C)$ reduces to the usual difference of geometric phases
acquired by the density operator in the closed case. In order to prove this,
consider the expansion of the vectors $|{\mathcal{D}}_{\alpha }\rangle
\rangle $ and $\langle \langle \mathcal{E}_{\alpha }|$ in a basis $%
\{I_{D},\Lambda _{i}|i=1,...,D^{2}-1\}$, where $I_{D}$ is the $D\times D$
identity matrix and $\Lambda _{i}$ are traceless Hermitian matrices, with ${%
\text{Tr}}(\Lambda _{i}\Lambda _{j})=\delta _{ij}$. Then, by using the
normalization condition $\langle \langle \mathcal{E}_{\alpha }|{\mathcal{D}}%
_{\alpha }\rangle \rangle =1$ and the matrix inner product $\langle \langle
u|v\rangle \rangle =(1/D)\,{\text{Tr}}(u^{\dagger }v)$~\cite{Horn:book}, we
obtain in the closed-case limit $|{\mathcal{D}}_{\alpha }\rangle \rangle
\rightarrow \sqrt{D}|\psi _{m}\rangle \langle \psi _{n}|$ [equivalently, $%
\langle \langle \mathcal{E}_{\alpha }|\rightarrow \sqrt{D}|\psi _{n}\rangle
\langle \psi _{m}|$], with $\{|\psi _{m}\rangle \}$ denoting a set of
normalized eigenstates of the Hamiltonian operator. Therefore, Eq.~(\ref%
{gp_abelian}) yields $\gamma _{\alpha }\rightarrow \gamma _{m}^{{\text{%
{\tiny {closed}}}}}-\gamma _{n}^{{\text{{\tiny {closed}}}}}$, which is
exactly the difference of phases acquired by the density matrix in closed
systems. Note that only phase \emph{differences} are experimentally
observable, so that the fact that our expression for the geometric phase
involves phase differences, rather than an absolute phase, is natural. As
mentioned above, this is an important aspect in which our expressions differ from
the ones derived using non-Hermitian Hamiltonians \cite%
{Garrison:88,Dattoli:90}.
\end{itemize}

\subsection{The non-Abelian Case: Generalized Holonomic Connection}

Let us now generalize these considerations to \emph{degenerate}
one-dimensional Jordan blocks, whence the geometric phase becomes
non-Abelian. From Eq.~(\ref{pgen}) we obtain 
\begin{equation}
{\dot{p}}_{\alpha }=-\sum_{\beta \,|\,\lambda _{\beta }=\lambda _{\alpha
}}p_{\beta }\langle \langle \mathcal{E}_{\alpha }|{\dot{\mathcal{D}}}_{\beta
}\rangle \rangle .  \label{p1D}
\end{equation}
Each decoupled subspace is associated with a different value of $\lambda
_{\alpha }$, and is spanned by the set $\{|\mathcal{D}_{\beta }(\vec{R}%
)\rangle \rangle \,|\,\lambda _{\beta }=\lambda _{\alpha }\}$. Then,
enumerating this list for each decoupled subspace (denoted by $\lambda
_{\alpha }$) as $|\mathcal{D}_{\lambda _{\alpha }}^{(1)}\rangle \rangle ,|%
\mathcal{D}_{\lambda _{\alpha }}^{(2)}\rangle \rangle ,...,|\mathcal{D}%
_{\lambda _{\alpha }}^{(G)}\rangle \rangle $ (the left basis $\{\langle
\langle \mathcal{E}_{\alpha }(\vec{R})|\}$ is similarly enumerated), with $G$
the degeneracy (dimension of the decoupled subspace), we have that ${\dot{p}}%
_{\lambda _{\alpha }}^{(i)}=-\sum_{j=1}^{G}p_{\lambda _{\alpha
}}^{(j)}\langle \langle \mathcal{E}_{\lambda _{\alpha }}^{(i)}|{\dot{%
\mathcal{D}}}_{\lambda _{\alpha }}^{(j)}\rangle \rangle $. Writing this
equation in a vector notation, we obtain 
\begin{equation}
{\dot{\mathbf{P}}}_{\lambda _{\alpha }}=-({\vec{A}}_{\lambda _{\alpha
}}\cdot {\dot{\vec{R}}})\mathbf{P}_{\lambda _{\alpha }},  \label{pvec}
\end{equation}%
where $\mathbf{P}_{\lambda _{\alpha }}=\left( p_{\lambda _{\alpha
}}^{(1)},\cdots ,p_{\lambda _{\alpha }}^{(G)}\right) ^{t}$ is a vector in
Hilbert-Schmidt space (superscript $t$ denotes transposition) and 
\begin{equation}
{\vec{A}}_{\lambda _{\alpha }}=\left( 
\begin{array}{cccc}
\langle \langle \mathcal{E}_{\lambda _{\alpha }}^{(1)}|\vec{\bigtriangledown}%
|{\mathcal{D}}_{\lambda _{\alpha }}^{(1)}\rangle \rangle & \cdots & \langle
\langle \mathcal{E}_{\lambda _{\alpha }}^{(1)}|\vec{\bigtriangledown}|{%
\mathcal{D}}_{\lambda _{\alpha }}^{(G)}\rangle \rangle &  \\ 
\vdots & \ddots & \vdots &  \\ 
\langle \langle \mathcal{E}_{\lambda _{\alpha }}^{(G)}|\vec{\bigtriangledown}%
|{\mathcal{D}}_{\lambda _{\alpha }}^{(1)}\rangle \rangle & \cdots & \langle
\langle \mathcal{E}_{\lambda _{\alpha }}^{(G)}|\vec{\bigtriangledown}|{%
\mathcal{D}}_{\lambda _{\alpha }}^{(G)}\rangle \rangle & 
\end{array}%
\right) .  \label{A}
\end{equation}%
Note that each element of ${\vec{A}}$ is a vector in parameter space (we use
bold-face and arrow superscripts to denote vectors in Hilbert-Schmidt space
and parameter space, respectively). The non-Abelian geometric phase in a
cyclic evolution associated with a degenerate level $\lambda _{\alpha }$ is
determined by the solution of Eq.~(\ref{pvec}), which is formally provided
by $\mathbf{P}_{\lambda _{\alpha }}(C)=\mathcal{U}\mathbf{\ P}_{\lambda
_{\alpha }}(0)$, where 
\begin{equation}
\mathcal{U}=\mathcal{P}\,e^{-\oint_{C}{\vec{A}}_{\lambda _{\alpha }}\cdot d{%
\vec{R}}}  \label{GP_1D}
\end{equation}%
is the corresponding Wilson loop, and $\mathcal{P}$ denotes path-ordering.

Equations~(\ref{A}) and (\ref{GP_1D}) constitute our second main result.
They are the generalization of the concept of non-Abelian geometric phases
to the open systems case. In particular, the matrix ${\vec{A}}_{\lambda
_{\alpha }}$ [Eq.~(\ref{A})] naturally generalizes the Wilczek-Zee gauge
potential~\cite{Wilczek:84}, also known as the holonomic connection.

A non-Abelian geometric phase will also appear in the case of
multi-dimensional Jordan blocks. However, in this case, it is not possible
to obtain a general analytical solution due to the presence of the term $%
p_{\alpha }^{(i+1)}$ in Eq.~(\ref{pgen}). One should then solve Eq.~(\ref%
{pgen}) on a case by case basis for all pairs $(\alpha ,i)$. This yields a
set of coupled differential equations in a ladder structure.

The geometric character of the non-Abelian geometric phase is evident from
the expression (\ref{GP_1D}) for the Wilson loop operator (it depends only
on the path and not on its parametrization). The closed system limit is
obtained in a manner exactly analogous to the proof above for the Abelian
case. What is left, therefore, in order to demonstrate that our expressions
for the generalized non-Abelian geometric phase have the desired properties,
is a proof of gauge invariance of the eigenvalues of the Wilson loop (this
is the same as in the closed systems case \cite{Wilczek:84}, where the
Wilson loop itself, without taking the trace out, is not gauge invariant).
We consider this issue next.

\subsection{Gauge invariance of the non-Abelian geometric phase in open
systems}

Let us show that the geometric phase derived in Eq.~(\ref{GP_1D}) is
invariant under gauge transformations. First, we can rewrite Eq. (\ref{pvec}%
) in the form 
\begin{equation}
{\dot{p}}_{\lambda _{\alpha }}^{(i)}=-\sum_{j=1}^{G}A_{\lambda _{\alpha
}}^{(ij)}p_{\lambda _{\alpha }}^{(j)}, 
\end{equation}%
where 
\begin{equation}
{A}_{\lambda _{\alpha }}^{(ij)}\equiv \langle \langle \mathcal{E}_{\lambda
_{\alpha }}^{(i)}|{\dot{\mathcal{D}}}_{\lambda _{\alpha }}^{(j)}\rangle
\rangle =\sum_{\mu }({A}_{\lambda _{\alpha }}^{(ij)})_{\mu }\frac{\partial
x_{\mu }(t)}{\partial t}=\overrightarrow{({A}_{\lambda _{\alpha }}^{(ij)})}%
\cdot {\dot{\vec{R}}},  \label{eq:A}
\end{equation}%
where we have introduced the explicit dependence on $\vec{R}%
(t)=(x_{1}(t),...,x_{n}(t))$, and 
\begin{equation}
({A}_{\lambda _{\alpha }}^{(ij)})_{\mu }\equiv \langle \langle \mathcal{E}%
_{\lambda _{\alpha }}^{(i)}|\partial _{\mu }|{\mathcal{D}}_{\lambda _{\alpha
}}^{(j)}\rangle \rangle .
\end{equation}%
Let us now apply a local gauge transformation $\Omega ({\vec{R}}(t))$ on the
right eigenvectors, 
\begin{equation}
|{\mathcal{D}}_{\lambda _{\alpha }}^{\prime (a)}\rangle \rangle
=\sum_{d}\Omega _{ad}|{\mathcal{D}}_{\lambda _{\alpha }}^{(d)}\rangle
\rangle ,
\end{equation}%
where $\Omega $ is an arbitrary complex matrix. Here $\Omega $ is taken as a
function of ${\vec{R}}(t)$ instead of $t$ because the basis vectors depend
on $t$ only through the parameters ${\vec{R}}(t)$. The normalization
constraint $\langle \langle \mathcal{E}_{\lambda _{\alpha }}^{\prime }|{%
\mathcal{D}}_{\lambda _{\alpha }}^{\prime }\rangle \rangle =1$ implies 
\begin{equation}
\langle \langle \mathcal{E}_{\lambda _{\alpha }}^{\prime
(b)}|=\sum_{c}\langle \langle \mathcal{E}_{\lambda _{\alpha }}^{(c)}|\Omega
_{cb}^{-1}.  \label{gt_vec}
\end{equation}%
The gauge transformation of ${A}_{\lambda _{\alpha }}^{(ba)}\equiv \langle
\langle \mathcal{E}_{\lambda _{\alpha }}^{(b)}|{\dot{\mathcal{D}}}_{\lambda
_{\alpha }}^{(a)}\rangle \rangle $, is then given by 
\begin{eqnarray}
{A}_{\lambda _{\alpha }}^{(ba)\prime } &=&\langle \langle \mathcal{E}%
_{\lambda _{\alpha }}^{\prime (b)}|{\dot{\mathcal{D}}}_{\lambda _{\alpha
}}^{\prime (a)}\rangle \rangle  \nonumber \\
&=&\sum_{cd}\Omega _{ad}(A_{\lambda _{\alpha }}^{t})_{dc}\Omega _{cb}^{-1}+%
\dot{\Omega}_{ad}\delta _{cd}\Omega _{cb}^{-1}  \nonumber \\
&=&(\Omega A_{\lambda _{\alpha }}^{t}\Omega ^{-1})_{ab}+(\dot{\Omega}\Omega
^{-1})_{ab},
\end{eqnarray}%
where superscript $t$ denotes transposition. Therefore 
\begin{equation}
A_{\lambda _{\alpha }}^{\prime t}=\Omega A_{\lambda _{\alpha }}^{t}\Omega
^{-1}+\dot{\Omega}\Omega ^{-1},  \label{ag}
\end{equation}%
which proves that $A_{\lambda _{\alpha }}^{t}$ transforms as a gauge
potential.

Now let us show that the Wilson loop has gauge-invariant eigenvalues (a
similar proof for the closed case can be found in Ref.~\cite{Alvarez:98}).
As a first step, consider the Wilson operator for an open curve 
\begin{widetext}
\begin{eqnarray}
\mathcal{U} &=&\mathcal{P}\exp \left( -\int_{{\vec{R}}(0)}^{{\vec{R}}(t)}{%
\vec{A}}\cdot d{\vec{R}}\right) =\mathcal{P}\exp \left(
-\int_{0}^{t}dt^{\prime }A_{\mu }(x)\frac{dx^{\mu }}{dt^{\prime }}\right)  
\nonumber \\
&=&1-\int_{0}^{t}dt_{1}A_{\mu _{1}}(t_{1})\frac{dx^{\mu _{1}}}{dt_{1}}%
+\int_{0}^{t}dt_{1}A_{\mu _{1}}(t_{1})\frac{dx^{\mu _{1}}}{dt_{1}}%
\int_{0}^{t_{1}}dt_{2}A_{\mu _{2}}(t_{2})\frac{dx^{\mu _{2}}}{dt_{2}}+... \, ,
\label{wo}
\end{eqnarray}%
where 
repeated indices are summed over,
and we suppress the subscript $\lambda _{\alpha }$ for 
notational simplicity. The 
transposed Wilson operator then yields
\begin{eqnarray}
\mathcal{U}^t &=& 1 - \int_{0}^{t} dt_1 A^t_{\mu_1}(t_1) \frac{dx^{\mu_1}}{%
dt_1} + \int_{0}^{t} dt_1 \int_{0}^{t_1} dt_2 A^t_{\mu_2}(t_2) \frac{%
dx^{\mu_2}}{dt_2} A^t_{\mu_1}(t_1) \frac{dx^{\mu_1}}{dt_1} + ...  \label{wot}
\end{eqnarray}
\end{widetext}
Note the inversion of the order of the operators due to the
transposition. 
Therefore, the transposed Wilson operator $W\equiv \mathcal{U}^{t}$ obeys
the differential equation 
\begin{equation}
\frac{dW}{dt}+WA_{\mu }^{t}\frac{dx^{\mu }}{dt}=0 . \label{woem2}
\end{equation}%
We can determine the gauge transformation of $W$ by imposing gauge
invariance of Eq.~(\ref{woem2}). After a gauge transformation, Eq.~(\ref%
{woem2}) reads: 
\begin{equation}
\frac{dW^{\prime }}{dt}+W^{\prime }A_{\mu }^{\prime t}\frac{dx^{\mu }}{dt}=0, 
\label{woemg}
\end{equation}%
where primes indicate gauge-transformed operators. Note that $A_{\mu }^{t}$,
which is given by 
\begin{equation}
(A_{\mu }^{t})_{ab}=\langle \langle \mathcal{E}_{\lambda _{\alpha
}}^{(b)}|\partial _{\mu }|{\mathcal{D}}_{\lambda _{\alpha }}^{(a)}\rangle
\rangle ,  \label{amu}
\end{equation}%
transforms, according to Eq.~(\ref{ag}), under a gauge transformation as 
\begin{equation}
A_{\mu }^{\prime t}=\Omega A_{\mu }^{t}\Omega ^{-1}+(\partial _{\mu }{\Omega 
})\Omega ^{-1}.  \label{amug}
\end{equation}%
Then, using Eq.~(\ref{amug}), we obtain 
\begin{equation}
\left( \frac{d(W^{\prime }\Omega )}{dt}+(W^{\prime }\Omega )A_{\mu }^{t}%
\frac{dx^{\mu }}{dt}\right) \Omega ^{-1}=0  . \label{woemgdev}
\end{equation}
Since $\Omega $ is arbitrary it follows from Eq.~(\ref{woemgdev}) that gauge
invariance of the equation of motion implies 
\begin{equation}
W\rightarrow W^{\prime }={\tilde{\Omega}}W\Omega ^{-1},  \label{W'}
\end{equation}%
where $\Omega ^{-1}=\Omega ^{-1}(x^{\mu }(t))$ and ${\tilde{\Omega}}$ is
independent of $x^{\mu }(t)$.

The gauge transformation of a product of paths allows us to further restrict ${%
\tilde{\Omega}}$:\ we can show that ${\tilde{\Omega}}=\Omega ^{-1}(x_{0})$,
where here and below $x_{i}\equiv x^{\mu }(t_{i})$ [$x^{\mu }(t_{0})$ is the
initial position $\vec{R}(0))$]. To see this, consider an open curve $\Gamma
=\Gamma _{1}+\Gamma _{2}$, where $\Gamma _{1}$ is a continuous curve in the
interval $[x_{0},x_{a}]$ and $\Gamma _{2}$ is a continuous curve in the
interval $[x_{a},x_{b}]$. The transposed Wilson operators $W(\Gamma _{1})$
and $W(\Gamma _{2})$ associated with these curves are, according to Eq.~(\ref%
{wot}), given by 
\begin{widetext}
\begin{eqnarray}
W(\Gamma_1) &=& 1 - \int_{0}^{t_a} dt_1 A^t_{\mu_1}(t_1) \frac{dx^{\mu_1}}{%
dt_1} + \int_{0}^{t_a} dt_1 \int_{0}^{t_1} dt_2 A^t_{\mu_2}(t_2) \frac{%
dx^{\mu_2}}{dt_2} A^t_{\mu_1}(t_1) \frac{dx^{\mu_1}}{dt_1} + ...  \nonumber \\
W(\Gamma_2) &=& 1 - \int_{t_a}^{t_b} dt_1 A^t_{\mu_1}(t_1) \frac{dx^{\mu_1}}{%
dt_1} + \int_{t_a}^{t_b} dt_1 \int_{t_a}^{t_1} dt_2 A^t_{\mu_2}(t_2) \frac{%
dx^{\mu_2}}{dt_2} A^t_{\mu_1}(t_1) \frac{dx^{\mu_1}}{dt_1} + ...  
\label{wgdef}
\end{eqnarray}
\end{widetext}
Then, under gauge transformation, we have from Eq.~(\ref{W'}): 
\begin{eqnarray}
W(\Gamma _{1}) &\rightarrow &{\tilde{\Omega}}_{1}W(\Gamma _{1})\Omega
(x_{a})^{-1}  \nonumber \\
W(\Gamma _{2}) &\rightarrow &{\tilde{\Omega}}_{2}W(\Gamma _{2})\Omega
(x_{b})^{-1}  \nonumber \\
W(\Gamma ) &\rightarrow &{\tilde{\Omega}}W(\Gamma )\Omega (x_{b})^{-1}.
\label{wgamma}
\end{eqnarray}%
Then, by applying a gauge transformation on $W(\Gamma )=W(\Gamma
_{1})W(\Gamma _{2})$, we obtain 
\begin{equation}
{\tilde{\Omega}}W(\Gamma )\Omega (x_{b})^{-1}={\tilde{\Omega}}_{1}W(\Gamma
_{1})\Omega (x_{a})^{-1}{\tilde{\Omega}}_{2}W(\Gamma _{2})\Omega
(x_{b})^{-1}.
\end{equation}
This implies (i) ${\tilde{\Omega}}_{2}=\Omega (x_{a})$, i.e., ${\tilde{\Omega%
}}$ depends only on the initial position, and (ii) ${\tilde{\Omega}}={\tilde{%
\Omega}}_{1}$ (${\tilde{\Omega}}$ must be independent of the path index). We
can therefore write ${\tilde{\Omega}}=\Omega (x_{0})$.Thus, for a general
open curve $\Gamma $ in the interval $[x_{0},x_{a}]$, the transposed Wilson
operator transforms as $W\rightarrow W^{\prime }=\Omega (x_{0})W\Omega
^{-1}(x_{a})$. For a closed curve $C$, we then get $W\rightarrow W^{\prime
}=\Omega (x_{0})W\Omega ^{-1}(x_{0})$, or, expressed in terms of the Wilson
loop itself: 
\begin{equation}
\mathcal{U}(C)\rightarrow \mathcal{U}^{\prime }(C)=(\Omega ^{t})^{-1}(x_{0})%
\mathcal{U}(C)\Omega ^{t}(x_{0}).  \label{wc}
\end{equation}%
Thus, the Wilson loop transforms as a similarity transformation, and
consequently its eigenvalues are gauge invariant, as desired. In particular,
this implies that the gauge-transformed amplitude vector $\mathbf{P}^{\prime
} _{\lambda _{\alpha }}\equiv (\Omega ^{t})^{-1}(x_{0})\mathbf{\ P}_{\lambda
_{\alpha }}$ obyes the same geometric phase transformation rule as $\mathbf{P%
}_{\lambda _{\alpha }}$: 
\begin{eqnarray}
\mathcal{U}^{\prime }\mathbf{P}^{\prime }_{\lambda _{\alpha }}(0) &=&(\Omega
^{t})^{-1}(x_{0})\mathcal{U}\Omega ^{t}(x_{0})(\Omega ^{t})^{-1}(x_{0})%
\mathbf{\ P}_{\lambda _{\alpha }}(0)  \nonumber \\
&=&(\Omega ^{t})^{-1}(x_{0})\mathbf{P}_{\lambda _{\alpha }}(C)=\mathbf{P}%
^{\prime }_{\lambda _{\alpha }}(C).
\end{eqnarray}

\section{Applications}
\label{apps}

\subsection{A Distinguished Time-Scale for Open System Geometric Phases}

As a first application of these general considerations we now show the
existence of a distinguished time-scale for the observation of open-system
geometric phases. To this end, it is convenient to express the variables in
terms of the dimensionless time $s=t/T$, where $T$ denotes the total
evolution time. Then, adiabatic dynamics in the interval $0\leq s\leq 1$
occurs if and only if the following time condition is satisfied: $T\gg
\max_{\alpha }\{T_{\alpha }^{c}\}$, where $T_{\alpha }^{c}$ denotes the 
\emph{crossover time} for the Jordan block $J_{\alpha }$~\cite%
{SarandyLidar:04}. For the particular case of one-dimensional blocks, we
have~\cite{SarandyLidar:04,SarandyLidar:05} 
\begin{eqnarray}
T_{\alpha }^{c} &=& \max_{0\leq s\leq 1} \left| \,\sum_{\beta \neq \alpha }
[ Q_{\beta \alpha}(0)-Q_{\beta \alpha}(s)\,e^{T\,\Omega _{\beta \alpha}(s)}
\right.  \nonumber \\
&&\left. +\int_{0}^{s}ds^{\prime }\,e^{T\,\Omega _{\beta \alpha }(s^{\prime
})}dQ_{\beta \alpha}(s^{\prime })/ds^{\prime } ] \right|,  \label{tcross}
\end{eqnarray}
where 
\begin{eqnarray}
\Omega _{\beta \alpha}(s) &=& \int_{0}^{s}\omega _{\beta \alpha }(s^{\prime
})\,ds^{\prime }  \nonumber \\
\omega _{\beta \alpha }(s) &=& \gamma _{\beta }(s)-\gamma _{\alpha }(s)
\end{eqnarray}
is the gap between Jordan eigenvalues, 
\begin{equation}
V_{\beta \alpha }(s)=p_{\beta }(s)\,\langle \langle \mathcal{E_{\alpha }}%
(s)| \frac{d\mathcal{L}(s)}{ds}|\mathcal{D_{\beta }}(s)\rangle \rangle
\end{equation}
are the matrix elements of the time-derivative of the Lindblad
superoperator, and 
\begin{equation}
Q_{\beta \alpha}(s) \equiv V_{\beta \alpha}(s)/\omega _{\beta \alpha
}^{2}(s).
\end{equation}
Note that a quantity analogous to $Q_{\beta \alpha}$, namely the
time-derivate of the Hamiltonian divided by the square of the spectral gap,
appears in the standard condition for adiabaticity in closed systems \cite%
{SarandyLidar:04}. In the expression for $V_{\beta \alpha }(s)$, upper
indices in $p_{\beta }^{(j)}(s)$ and in the basis vectors $\{|{\mathcal{D}%
_{\beta }^{(j)}(s)\rangle \rangle }\}$ and $\{\langle \langle \mathcal{E}%
_{\alpha }^{(i)}(t)|\}$ were removed because the Jordan blocks are
one-dimensional. The crossover time $T_{\alpha }^{c}$ provides a decoupling
timescale for each Jordan block: \emph{provided $T \gg T_{\alpha }^{c}$ the
Jordan block $J_{\alpha }$ is adiabatically decoupled from all other blocks
associated to a different eigenvalue}. The general expression for $T_{\alpha
}^{c}$ in the case of multi-dimensional Jordan blocks, as well as a more
detailed discussion of its meaning, are given in Refs.~\cite%
{SarandyLidar:04,SarandyLidar:05}.

Now, the important observation is that the decoupling time-scale is \emph{%
finite} due to the presence of complex exponentials in $T_{\alpha }^{c}$~%
\cite{SarandyLidar:04}, which have real and imaginary parts in the case of
open systems. Therefore, since the geometric phases are defined in the
adiabatic regime, they will only be observable during the finite time in
which the Jordan blocks are decoupled. This fact implies the existence of a
distinguished, finite time-scale for geometric phases in open systems. Such
a time-scale was noted in Ref.~\cite{Whitney:03} in the context of a
specific example, namely the case of a spin-1/2 particle in a magnetic
field. Finite adiabaticity time-scales have been revealed as a general
property of open systems~\cite{SarandyLidar:04,Thunstrom:05,Rezakhani:05_2},
a fact which has also been observed in adiabatic QC, both theoretically~\cite%
{SarandyLidar:05} and experimentally~\cite{Steffen:03}. Physically, the
reason for this phenomenon is the broadening of the system energy levels due
to the presence of a dense spectrum of bath energy levels, until the
broadened system energies overlap. When this happens different eigenspaces
may no longer be decoupled (provided there are no selection rules preventing
the coupling), and the adiabatic approximation breaks down. In the case of
static Hamiltonians this is known as quantum diffusion~\cite{Dykman:78}.

\subsection{Spin $1/2$ in a time-dependent magnetic field under decoherence}

As an illustration of the general theory presented above, let us consider
the canonical example of a spin-$1/2$ in a time-dependent magnetic field,
originally considered by Berry in the context of closed quantum systems~\cite%
{Berry:84}. The renormalized Hamiltonian of the system is given by $H(\vec{B}%
)=-\mu \vec{S}\cdot \vec{B}$, where $\vec{S}=(1/2)(\sigma ^{x},\sigma
^{y},\sigma ^{z})$ is the spin operator, with $\sigma ^{i}$ ($i=x,y,z$)
denoting the Pauli matrices, $\vec{B}(t)=(B_{x}(t),B_{y}(t),B_{z}(t))$ is a
time-dependent magnetic field (including the Lamb shift correction \cite
{Alicki:87,Lidar:CP01}), and $\mu $ is a constant. A standard evaluation of the
geometric phase in this case yields $\gamma _{\pm }^{\mathrm{closed}}(C)=\pm
\Omega (C)/2$, where $\gamma _{\pm }^{\mathrm{closed}}(C)$ are the geometric
phases associated with the energy levels $E_{\pm }=\pm (\mu /2)B$, with $B=|%
\vec{B}(t)|$, and $\Omega (C)$ being the solid angle subtended by the closed
curve $C$ traversed by the magnetic field in parameter space.

In the weak-coupling regime, it is common to consider decoherence in the
eigenbasis of the renormalized system Hamiltonian \cite%
{ALZ:05,Childs:02,Kaminsky:04}. Let us now analyze the effects of
decoherence in this basis, assuming that open-systems dynamics is described
by the master equation~(\ref{eq:t-Lind2}). We consider two important sources
of decoherence, namely, dephasing and spontaneous emission in the
eigenenergy basis. The Lindblad operators modelling these processes are
given, respectively, by $\Gamma _{z}=\beta _{z}W(\vec{B})\sigma
_{z}W^{\dagger }(\vec{B})$ and $\Gamma _{-}=\beta _{-}W(\vec{B})\sigma
_{-}W^{\dagger }(\vec{B})$ ($\sigma _{-}|1\rangle =2|0\rangle ,\sigma
_{-}|0\rangle =0$), where $\beta _{z}$ and $\beta _{-}$ are the error
probabilities per unit time and $W(\vec{B})$ is the unitary matrix which
diagonalizes $H(\vec{B})$. The Lindblad superoperator is then given by $%
\mathcal{L}(\vec{B})=\mathcal{H}(\vec{B})+\mathcal{R}(\vec{B})$, where $%
\mathcal{H}(\vec{B})$ is the Hamiltonian superoperator [obtained from $H(%
\vec{B})$] and $\mathcal{R}(\vec{B})$ is the superoperator containing the
decoherence contribution [obtained from $\Gamma _{z}$ and $\Gamma _{-}$]. In
this case explicit calculation reveals that (i) $\mathcal{H}$ and $\mathcal{R%
}$ are diagonalizable, (ii) $\left[ \mathcal{H},\mathcal{\ R}\right] =0$.
Hence $\mathcal{L}$, $\mathcal{H}$ and $\mathcal{R}$ have a common
eigenstate basis, and in particular it follows that $\mathcal{L}$ has only
one-dimensional Jordan blocks, whence it is diagonalizable. Thus, bearing in
mind that the eigenstate basis for $\mathcal{H}(\vec{B})$ is independent of $%
\beta _{z}$ and $\beta _{-}$, it follows that the eigenstate basis for $%
\mathcal{L}(\vec{B})$ is also independent of $\beta _{z}$ and $\beta _{-}$.
This implies that \emph{the adiabatic geometric phases, which can be
computed here from Eq.~(\ref{gp_abelian}), are robust against dephasing} ($%
\Gamma _{z}$) \emph{and spontaneous emission} ($\Gamma _{-}$). In this case,
integration in parameter space, which is the relevant space for adiabatic
geometric phases, is not affected by decoherence. Nevertheless, the
adiabaticity crossover time $T_{\alpha }^{c}$ does depend on $\beta _{z}$
and $\beta _{-}$ through the eigenvalues of $\mathcal{L}$.

The robustness against dephasing is in agreement with Ref.~\cite{Carollo:03}%
, but obtained here in a totally different framework. A microscopic
derivation of the geometric phase for a spin-1/2 in a magnetic field was
developed in Ref.~\cite{Whitney:05}, with no robustness against dephasing
detected. However, 
note that the robustness of the geometric phase obviously depends on the 
basis in which the environment acts. As is clear from  Eq.~(4) of
Ref.~\cite{Whitney:05}, Whitney et al. consider dephasing in a
\emph{fixed} (time-independent) basis, where robustness is absent. 
In this basis, our approach is in agreement with 
this lack of robustness. However, in the example we discuss here,
we consider decoherence in the \emph{instantaneous} eigenenergy basis. In the 
weak-coupling regime, one should consider decoherence in the eigenbasis of the 
system Hamiltonian (see, e.g., Refs.~\cite{Childs:02,Kaminsky:04}). 
This follows from the fundamental Davies derivation of the quantum Markovian master
equation~\cite{Davies:78}, which was recently reviewed and generalized
in a context relevant to ours in Ref.~\cite{ALZ:05}, and which shows that in
the Markovian limit time-dependent system Hamiltonians are always
coupled to the Lindblad operators. This difference in basis explains
the apparent disparity between Ref.~\cite{Whitney:05} and our result. Then, in the
instantaneous eigenbasis, we obtain robustness 
against both dephasing and spontaneous emission for adiabatic evolution as a 
simple consequence of the commutation relation between the Hamiltonian superoperator ${\cal H}$ and the corresponding decoherence superoperators
${\cal R}$. In fact, the commutation between ${\cal H}$ and ${\cal R}$ provides a general 
sufficient condition for robustness of adiabatic geometric phases against ${\cal R}$.
Note also that the non-adiabatic geometric
phase is usually affected by corrections due to the system-bath interaction,
in particular in the case of spontaneous emission~\cite{Carollo:03}.
Remarkably, by imposing adiabaticity on the open system, robustness of the
geometric phase against this decoherence process is obtained.

\begin{figure}[ht]
\centering {\includegraphics[angle=0,scale=0.35]{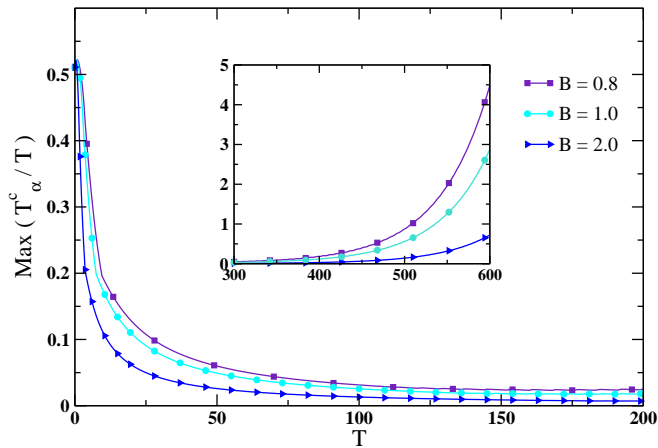}}
\caption{Maximum value of the ratio $T_\protect\alpha^c/T$ of the crossover
time to the total evolution time, taken over all the Jordan blocks, as a
function of $T$ for a spin-$1/2$ particle in a magnetic field undergoing
dephasing. Parameter values are given in the text. The adiabatic interval
requires $T_\protect\alpha^c \ll T$. Observe that the stronger the magnetic
field (at fixed decoherence rate), the better the adiabatic approximation,
and hence the longer we can observe the geometric phase. }
\label{f1}
\end{figure}

We stress that, to the best of our knowledge, the approach presented here
for dealing with geometric phases is the first to predict robustness against
both dephasing and spontaneous emission for a spin-1/2. We expect that such
a robustness will serve as a useful protection mechanism in holonomic QC
(see, e.g., Ref.~\cite{Guridi:05} for difficulties in the correction of
spontaneous emission). The robustness is reminiscent of the emergence of a
decoherence-free subspace (DFS) \cite{LidarWhaley:03}, but unlike the
symmetry-driven appearance of the latter, here the robustness is due to a,
fundamentally different, adiabatic mechanism. Related observations were
made, using very different methods, in Ref.~\cite{Carollo:06}, for a system
in a DFS, coupled to a cyclicly evolving reservoir.

\begin{figure}[ht]
\centering {\includegraphics[angle=0,scale=0.35]{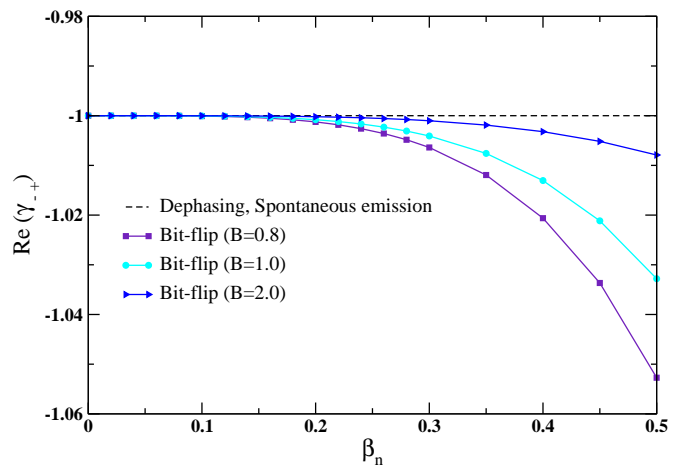}}
\caption{Geometric phase (in units of $\protect\pi$) of a spin-1/2 in a
magnetic field, as a function of the decoherence strengths $\protect\beta%
_{n} $ ($n$ is chosen as $z$, $-$, or $x$ depending on the decoherence
process, as indicated in the legend). The magnetic field is applied in a
spherically symmetric configuration whose parameters are chosen as in Fig.~%
\protect\ref{f1}. The geometric phase $\protect\gamma_{-+}(C)$ plotted is
for the Jordan block associated -- in the limit of vanishing decoherence --
with the closed-system geometric phase difference $\protect\gamma_{-}^{%
\mathrm{closed}}(C)-\protect\gamma_{+}^{\mathrm{closed}}(C)$, which equals $%
-2\protect\pi[1-\cos{(\protect\theta=\protect\pi/3})]=-\protect\pi$.}
\label{f2}
\end{figure}

As for closed systems, $B$ plays an important role in setting the time
interval for the observation of the geometric phase. The reason is that the
Zeeman effect further splits the system energy levels, thus postponing the
breakdown of adiabaticity due to overlap caused by environmentally induced
broadening. This behavior is illustrated in Fig.~\ref{f1} for the case of
dephasing, where we take $\beta_z=0.1$ (in units such that $\mu=1$) and use
the following spherically symmetric configuration for the magnetic field: $%
B_x (s) = B \cos (2 \pi s) \sin \theta$, $B_y (s) = B \sin (2 \pi s) \sin
\theta$, and $B_z = B \cos \theta$, with $\theta$ denoting the azimuthal
angle, set at $\pi/3$. The initial state of the system is chosen to be an
equal superposition of energy eigenstates. Due to the commutation relation $[%
\mathcal{H},\mathcal{R}]=0$, we find that the four Jordan blocks are
associated with the set of Hilbert-Schmidt space vectors $%
\{|\psi_m\rangle\langle \psi_n|\}$ $(m,n=1,2)$, where $\{|\psi_m\rangle\}$
are the normalized eigenstates of the system Hamiltonian. Then, direct
computation of $\gamma_{\alpha}$ from Eq.~(\ref{gp_abelian}) (for $s=1$)
yields $\gamma_{\alpha}=\pm 2\pi \cos{\theta}$ for the vectors $%
|\psi_m\rangle\langle \psi_n|$ with $m\ne n$. Up to an unimportant $2\pi$
factor, these are exactly the differences of geometric phases $\pm
2\pi(1-\cos{\theta})$ appearing in the density operator for these states in
the closed case. As suggested from the analytical treatment above, similar
results hold for spontaneous emission.

It should be noticed, however, that robustness is, naturally, not universal;
e.g., it does not hold for the bit-flip channel $\Gamma_x(\vec{B}) = \beta_x
W(\vec{B}) \sigma_x W(\vec{B})^{\dagger}$, since $\Gamma_x$ does not commute
with the Hamiltonian superoperator. This result is illustrated in Fig.~\ref%
{f2}, where it is shown that the real part of the geometric phase under
bit-flip is slightly affected by the decoherence process. This is in
contrast with the robustness of dephasing and spontaneous emission. The
corresponding imaginary part of the geometric phase (not shown) has
negligible variation, which means that the visibility of the phase is not
significantly affected by the bit-flip channel for the decoherence strengths
considered in the plot.

\section{Conclusions}
\label{conc}

We have introduced a general framework for geometric phases in open systems
undergoing cyclic adiabatic evolution. Expressions which naturally
generalize the familiar closed systems' (Abelian) Berry phase and
(non-Abelian) Wilczek-Zee gauge potential and Wilson loop were derived, and
their gauge invariance proven. An important feature of our approach is the
existence of a distinguished time-scale for the observation of the adiabatic
geometric phase. This property imposes time constraints on realistic schemes
for holonomic QC based on adiabatic phases. Remarkably, robustness against
dephasing and spontaneous emission was found for the geometric phase
acquired by a spin-1/2 in a magnetic field. This robustness is, however, not
universal; e.g., it does not hold for the bit-flip channel, since the latter
does not commute with the Hamiltonian superoperator. The development of
methods for overcoming decoherence affecting
geometric phases is
therefore of significant interest \cite{Guridi:05,WuZanardiLidar:05}.

\section*{Acknowledgements}

We gratefully acknowledge financial support from CNPq and FAPESP (to M.S.S.)
and the Sloan Foundation (to D.A.L.). We also thank Dr. P. Zanardi and Dr.
L. A. Ferreira for their valuable comments.

\begin{appendix}
\section{}
\label{appA} 

We define the right $\{|\mathcal{D}_{\beta }^{(j)}[\vec{R}(t)]\rangle
\rangle \}$ and left $\{\langle \langle \mathcal{E}_{\alpha }^{(i)}[\vec{R}%
(t)]|\}$ basis vectors associated with $\mathcal{L}[\vec{R}(t)]$ and 
prove their bi-orthogonality and completeness. Instantaneous right and left
eigenstates of a general time-dependent superoperator $\mathcal{L}(t)$ are 
defined by 
\begin{eqnarray}
\mathcal{L}(t)\,|\mathcal{P}_{\alpha }(t)\rangle \rangle  &=&\lambda
_{\alpha }(t)\,|\mathcal{P}_{\alpha }(t)\rangle \rangle ,  \label{rleb0} \\
\langle \langle \mathcal{Q}_{\alpha }(t)|\,\mathcal{L}(t) &=&\langle \langle 
\mathcal{Q}_{\alpha }(t)|\,\lambda _{\alpha }(t),  \label{rleb}
\end{eqnarray}%
where possible degeneracies correspond to $\lambda _{\alpha }=\lambda
_{\beta }$, with $\alpha \neq \beta $. In other words, we reserve a
different index $\alpha $ for each independent eigenvector since each
eigenvector is in a distinct Jordan block. It follows from Eqs.~(\ref{rleb0}%
) and~(\ref{rleb}) that, on the one hand%
\begin{equation}
\langle \langle \mathcal{Q}_{\beta }(t)|(\mathcal{L}(t)\,|\mathcal{P}%
_{\alpha }(t)\rangle \rangle )=\lambda _{\alpha }(t)\langle \langle \mathcal{%
Q}_{\beta }(t)\,|\mathcal{P}_{\alpha }(t)\rangle \rangle ,
\end{equation}%
while on the other hand this equals%
\begin{equation}
(\langle \langle \mathcal{Q}_{\beta }(t)|\,\mathcal{L}(t))\,|\mathcal{P}%
_{\alpha }(t)\rangle \rangle =\langle \langle \mathcal{Q}_{\beta }(t)\,|%
\mathcal{P}_{\alpha }(t)\rangle \rangle \,\lambda _{\beta }(t).
\end{equation}%
Therefore for $\lambda _{\alpha }\neq \lambda _{\beta }$, we have $\langle
\langle \mathcal{Q}_{\alpha }(t)|\mathcal{P}_{\beta }(t)\rangle \rangle =0$.

The left and right eigenstates can be easily identified when the Lindblad
superoperator is in the Jordan form $\mathcal{L}_{\mathrm{J}}(t)=S^{-1}(t)\mathcal{L}S(t)$. 
Denoting $|\mathcal{P}_{\alpha }(t)\rangle \rangle
_{J}=S^{-1}(t)\,|\mathcal{P}_{\alpha }(t)\rangle \rangle $, i.e., the right
\textquotedblleft Jordan basis\textquotedblright\ (note the $J$ subscript)\
eigenstate of $\mathcal{L}_{\mathrm{J}}(t)$ associated to a Jordan block $%
J_{\alpha }$, then Eq.~(\ref{rleb0}) implies that $|\mathcal{P}_{\alpha
}(t)\rangle \rangle _{J}$ is time-independent and, after normalization, is
given by 
\begin{equation}
\left. |\mathcal{P}_{\alpha }\rangle \rangle _{J}\frac{{}}{{}}\right\vert
_{J_{\alpha }}=\left( \frac{{}}{{}}1,0,0,\,.\,.\,.\,,0\frac{{}}{{}}\right)
^{t},  \label{pj}
\end{equation}%
where only the vector components associated to the Jordan block $J_{\alpha }$
are shown, with all the others vanishing. In order to have a complete basis
we shall define new states, which will be chosen so that they preserve the
block structure of $\mathcal{L}_{\mathrm{J}}(t)$. A suitable set of
additional vectors is 
\begin{eqnarray}
\left. |\mathcal{D}_{\alpha }^{(1)}\rangle \rangle _{J}\frac{{}}{{}}%
\right\vert _{J_{\alpha }} &=&\left( \frac{{}}{{}}0,1,0,\,.\,.\,.\,,0\frac{{}%
}{{}}\right) ^{t},\,...\,,  \nonumber \\
\,\left. |\mathcal{D}_{\alpha }^{(n_{\alpha }-1)}\rangle \rangle _{J}\frac{{}%
}{{}}\right\vert _{J_{\alpha }} &=&\left( \frac{{}}{{}}0,,\,.\,.\,.\,,0,1%
\frac{{}}{{}}\right) ^{t},  \label{dj}
\end{eqnarray}%
where again all the components outside $J_{\alpha }$ are zero. This simple
vector structure allows for the derivation of the expression 
\begin{equation}
\mathcal{L}_{\mathrm{J}}(t)\,|\mathcal{D}_{\alpha }^{(j)}\rangle \rangle
_{J}=|\mathcal{D}_{\alpha }^{(j-1)}\rangle \rangle _{J}+\lambda _{\alpha
}(t)\,|\mathcal{D}_{\alpha }^{(j)}\rangle \rangle _{J},  \label{ldj}
\end{equation}%
with $|\mathcal{D}_{\alpha }^{(0)}\rangle \rangle _{J}\equiv |\mathcal{P}%
_{\alpha }\rangle \rangle _{J}$ and \ $|\mathcal{D}_{\alpha }^{(-1)}\rangle
\rangle _{J}\equiv 0$. The set $\left\{ |\mathcal{D}_{\alpha }^{(j)}\rangle
\rangle _{J}\text{, \thinspace }\,j=0,...,(n_{\alpha }-1)\right\} $ can
immediately be related to a right vector basis for the original $\mathcal{L}%
(t)$ by means of the transformation $|\mathcal{D}_{\alpha }^{(j)}(t)\rangle
\rangle =S(t)\,|\mathcal{D}_{\alpha }^{(j)}\rangle \rangle _{J}$ which,
applied to Eq.~(\ref{ldj}), yields 
\begin{equation}
\mathcal{L}(t)\,|\mathcal{D}_{\alpha }^{(j)}(t)\rangle \rangle =|\mathcal{D}%
_{\alpha }^{(j-1)}(t)\rangle \rangle +\lambda _{\alpha }(t)\,|\mathcal{D}%
_{\alpha }^{(j)}(t)\rangle \rangle .  \label{ldo}
\end{equation}%
Equation~(\ref{ldo}) exhibits an important feature of the set $\left\{ |%
\mathcal{D}_{\beta }^{(j)}(t)\rangle \rangle \right\} $, namely, it implies
that Jordan blocks are invariant under the action of the Lindblad
superoperator, i.e., the index $\alpha$ denoting the Jordan block is preserved 
under $\cal L$.

An analogous procedure can be employed to define the left basis.
Denoting by $_{J}\langle \langle \mathcal{Q}_{\alpha }(t)|=\langle \langle 
\mathcal{Q}_{\alpha }(t)|S(t)$ the left eigenstate of $\mathcal{L}_{\mathrm{J%
}}(t)$ associated to a Jordan block $J_{\alpha }$, Eq.~(\ref{rleb}) leads to
the normalized left vector 
\begin{equation}
\left. _{J}\langle \langle \mathcal{Q}_{\alpha }|\frac{{}}{{}}\right\vert
_{J_{\alpha }}=\left( \frac{{}}{{}}0,\,.\,.\,.\,,0,1\frac{{}}{{}}\right) .
\label{qj}
\end{equation}%
The additional left vectors are defined as (note that these are just the
transpose of the right vectors in the Jordan basis) 
\begin{eqnarray}
\left. _{J}\langle \langle \mathcal{E}_{\alpha }^{(0)}|\frac{{}}{{}}%
\right\vert _{J_{\alpha }} &=&\left( \frac{{}}{{}}1,0,0,\,.\,.\,.\,,0\frac{{}%
}{{}}\right) ,  \nonumber \\
\vspace{0.1cm} &.\,\,.\,\,.&  \nonumber \\
\vspace{0.1cm}\left. _{J}\langle \langle \mathcal{E}_{\alpha }^{(n_{\alpha
}-2)}|\frac{{}}{{}}\right\vert _{J_{\alpha }} &=&\left( \frac{{}}{{}}%
0,\,.\,.\,.\,,0,1,0\frac{{}}{{}}\right) ,  \label{rj}
\end{eqnarray}%
which imply the following expression for the left basis vector $\langle
\langle \mathcal{E}_{\alpha }^{(i)}(t)|=\,_{J}\langle \langle \mathcal{E}%
_{\alpha }^{(i)}|\,S^{-1}(t)$ for $\mathcal{L}(t)$: 
\begin{equation}
\langle \langle \mathcal{E}_{\alpha }^{(i)}(t)|\,\mathcal{L}(t)=\langle
\langle \mathcal{E}_{\alpha }^{(i+1)}(t)|+\langle \langle \mathcal{E}%
_{\alpha }^{(i)}(t)|\,\lambda _{\alpha }(t),  \label{lro}
\end{equation}%
or, equivalently, 
\begin{equation}
\mathcal{L}(t)^{\dag }|\mathcal{E}_{\alpha }^{(i)}\rangle \rangle =\lambda
_{\alpha }^{\ast }(t)|\mathcal{E}_{\alpha }^{(i)}\rangle \rangle +|\mathcal{E%
}_{\alpha }^{(i+1)}\rangle \rangle \text{.}
\end{equation}%
Here we have used the notation $_{J}\langle \langle \mathcal{E}_{\alpha
}^{(n_{\alpha }-1)}|\equiv \,_{J}\langle \langle \mathcal{Q}_{\alpha }|$ and 
$_{J}\langle \langle \mathcal{E}_{\alpha }^{(n_{\alpha })}|\equiv 0$. 

We can now derive the orthogonality and completeness relations. First, the
left and right basis vectors are orthonormal: 
\begin{equation}
\langle \langle \mathcal{E}_{\alpha }^{(i)}(t)|\mathcal{D}_{\beta
}^{(j)}(t)\rangle \rangle =\,_{J}\langle \langle \mathcal{E}_{\alpha
}^{(i)}|S^{-1}(t)S(t)|\mathcal{D}_{\beta }^{(j)}\rangle \rangle
_{J}=\delta _{\alpha \beta }\delta ^{ij}.  \label{lrr}
\end{equation}%
Second, it is clear that (since it is a standard basis) the Jordan basis is
complete in the sense that $\sum_{\alpha ,\beta ;i,j}|\mathcal{D}_{\beta
}^{(j)}\rangle \rangle_{J}\, _{J}\langle \langle \mathcal{E}_{\alpha }^{(i)}|=I$.
Applying $S(t)$ to the left and $S^{-1}(t)$ to the right of this equation we
therefore find the completeness relation%
\begin{equation}
\sum_{\alpha ,\beta ;i,j}|\mathcal{D}_{\beta }^{(j)}(t)\rangle \rangle
\langle \langle \mathcal{E}_{\alpha }^{(i)}(t)|=I.
\end{equation}
As a final point of clarification, note that, even though in the stationary
Jordan basis left and right basis states coincide (up to transposition), this
is not the case in the time-dependent basis. This difference between left
and right vectors is due to the non-unitarity of the similarity matrix $%
S$. To see this, note that for a 1-dimensional Jordan block: $|\mathcal{D}%
_{\alpha }\rangle \rangle _{J}=|\mathcal{E}_{\alpha }\rangle \rangle _{J}$,
so instead of $\langle \langle \mathcal{E}_{\alpha }(t)|=\,_{J}\langle
\langle \mathcal{E}_{\alpha }|S^{-1}(t)$ we can write $\langle \langle 
\mathcal{E}_{\alpha }(t)|=\,_{J}\langle \langle \mathcal{D}_{\alpha
}|S^{-1}(t)$. Then $|\mathcal{E}_{\alpha }(t)\rangle \rangle =(S^{-1})^\dag (t)|%
\mathcal{D}_{\alpha }\rangle \rangle _{J}$, which does not equal $|\mathcal{D%
}_{\alpha }(t)\rangle \rangle =S(t)|\mathcal{D}_{\alpha }\rangle \rangle _{J}
$ since $S$ is not unitary.

\end{appendix}

\end{document}